\documentclass[aps,prd,twocolumn,preprintnumbers,
showpacs,floatfix]{revtex4}

\usepackage{bm}
\usepackage{epsfig}
\usepackage{graphics}

\newcommand{\eeq}{\end{equation}}
\newcommand{\beq}{\begin{equation}}
\newcommand{\ba}{\begin{array}}
\newcommand{\ea}{\end{array}}
\newcommand{\bea}{\begin{eqnarray}}
\newcommand{\eea}{\end{eqnarray}}
\newcommand{\vev}[1]{\langle #1\rangle}

\newcommand{\eps}{\epsilon}

\begin{document}

\preprint{UT-STPD-4/03}

\title{Leptogenesis through direct inflaton
decay to light particles}

\author{Thomas Dent}
\email{tdent@gen.auth.gr}
\author{George Lazarides}
\email{lazaride@eng.auth.gr}
\author{Roberto Ruiz de Austri}
\email{rruiz@gen.auth.gr}
\affiliation{Physics Division, School of
Technology, Aristotle University of
Thessaloniki, Thessaloniki 54124, Greece}

\date{\today}

\begin{abstract}
\noindent
We present a scenario of nonthermal
leptogenesis following supersymmetric hybrid
inflation, in the case where inflaton decay
to both heavy right handed neutrino and
${\rm SU}(2)_L$ triplet superfields is
kinematically disallowed. Lepton asymmetry
is generated through the decay of the
inflaton into light particles by the
interference of one-loop diagrams with right
handed neutrino and ${\rm SU}(2)_L$ triplet
exchange respectively. We require
superpotential couplings explicitly
violating a ${\rm U}(1)$ R-symmetry and
R-parity. However, the broken R-parity need
not have currently observable low-energy
signatures. Also, the lightest sparticle can
be stable. Some R-parity violating slepton
decays may, though, be detectable in the
future colliders. We take into account the
constraints from neutrino masses and mixing
and the preservation of the primordial lepton
asymmetry.
\end{abstract}

\maketitle

\section{Introduction}
\label{sec:intro}

\par
The standard model (SM) of electroweak and strong
interactions, despite its many successes, leaves
unanswered many questions in particle physics and
cosmology. In particular, it does not address the
following problems: the origin of electroweak
symmetry breaking and of the hierarchy between
the electroweak scale $M_W$ and the (reduced)
Planck scale $m_{\rm P}\simeq 2.44\times 10^{18}
~{\rm GeV}$; the origin and size of neutrino
($\nu$) masses and mixing; the cosmological
horizon and flatness problems and the origin of
density perturbations; the generation of the
observed baryon asymmetry of the universe (BAU).

\par
The problem of stabilizing the electroweak scale
relative to the fundamental scale $M_S$ (which we
take to be close to the Planck scale) is solved
by softly-broken supersymmetry (SUSY). For
definiteness, we take the case of gravity-mediated
SUSY-breaking, where the gravitino mass is of
order $1~{\rm TeV}$. In this framework, the vacuum
expectation values (VEVs) of the electroweak Higgs
superfields $h_1$, $h_2$ can be determined by
radiative corrections to the scalar potential
\cite{radcorr}. However, this explanation of the
electroweak symmetry-breaking requires a term
$\mu h_1h_2$ in the superpotential with
$\mu\sim 1~{\rm TeV}$, whereas a priori the
value of $\mu$ is expected to be of the order of
the fundamental scale. This $\mu$ problem of
the minimal SUSY standard model (MSSM) can
be addressed by imposing a symmetry that forbids
the above $\mu$ term, which is then broken in a
controlled fashion. A R-symmetry, broken by the
soft SUSY-breaking terms, can fulfill \cite{lr}
this role (see below).

\par
The smallness of neutrino masses is naturally
explained in a model with additional SM singlet
chiral fermions ($\nu^c$), the right handed
neutrinos (RHNs), since these can obtain heavy
Majorana masses ($\sim M_{\nu^c}$) and also
participate in Yukawa couplings with the
${\rm SU}(2)_L$ doublet neutrinos resulting in
Dirac neutrino masses ($\sim m_D$) after the
electroweak symmetry breaking. The resulting
seesaw mass matrix \cite{seesaw} has small
eigenvalues of order $m_D^2/M_{\nu^c}$,
which can be of the right order of magnitude
to explain atmospheric and solar neutrino
oscillations, while being consistent with
cosmological bounds on neutrino masses.
Another mechanism of inducing nonzero neutrino
masses is the introduction of heavy scalar
${\rm SU}(2)_L$ triplets $T$ with lepton
number $-2$, coupling to the ${\rm SU}(2)_L$
doublet lepton fields. After the breaking of
the electroweak symmetry, the neutral
components of these triplets can acquire
small VEVs inducing Majorana neutrino masses
\cite{triplet}. In general, these two
mechanisms for generating neutrino masses can
coexist.

\par
The cosmological horizon and flatness problems
are most elegantly solved by inflation, which
can also generate the primordial density
perturbations required for structure formation
in the universe \cite{llbook}. Moreover,
inflation, which can be easily incorporated in
realistic particle physics models, is strongly
favored by the recent data \cite{wmap} on the
angular power spectrum of the cosmic microwave
background radiation (CMBR).

\par
If neutrinos get mass either by coupling to
heavy SM singlet fermions or to heavy scalar
${\rm SU}(2)_L$ triplets, it may be possible to
generate \cite{lepto,tripletdecay} a primordial
lepton asymmetry in the out-of-equilibrium
decay of these heavy degrees of freedom, in
case they were thermally produced in the early
universe. This asymmetry is then reprocessed at
the electroweak phase transition to yield the
observed BAU. However, there is a tension
between correct neutrino masses and this
thermal leptogenesis scenario in SUSY models
because of the gravitino problem
\cite{khlopov,gravitino}. With a gravitino
mass of around $1~{\rm TeV}$, the reheat
temperature $T_{\rm rh}$ should not exceed
$10^{9}~{\rm GeV}$ since gravitinos produced
in thermal scattering processes would decay
late presumably into photons and photinos and,
if sufficiently numerous, interfere with the
successful predictions of standard big bang
nucleosynthesis. On the other hand, one also
requires that the heavy fields whose decay
creates lepton asymmetry be generated in
sufficient numbers. So, their masses should
not exceed $T_{\rm rh}$, which leads to
unacceptably large light neutrino masses.
However, this problem can be alleviated
\cite{pilaftsis,deg} by assuming that there
is some degree of degeneracy between the
relevant RHNs or heavy ${\rm SU}(2)_L$ triplets,
which enhances the generated lepton asymmetry,
and perhaps also that the branching ratio of
the gravitino decay into photons and photinos
is less than unity, which somewhat relaxes
\cite{gravitino} the gravitino constraint on
$T_{\rm rh}$.

\par
The tension between $\nu$ masses and the
gravitino problem can be more naturally
relaxed by considering nonthermal
leptogenesis \cite{inflepto} at reheating.
However, in existing scenarios
\cite{virtualtriplet,nonthtripletdec},
where the inflaton decays into RHN or
${\rm SU}(2)_L$ triplet superfields, this
still puts a restriction on the masses of
these particles: the decay products of the
inflaton must be lighter than half its mass
$m_{\rm inf}$. Lepton asymmetry is generated
in the subsequent decay of the RHN or
${\rm SU}(2)_L$ triplet superfields.

\par
In this work, however, we consider the
consequences of allowing all the RHN and/or
${\rm SU}(2)_L$ triplet superfields of the
model to be heavier than $m_{\rm inf}/2$
(see also Ref.~\cite{allahv}). Leptogenesis
could then occur only via the direct decay
of the inflaton to light particles
(see also Ref.~\cite{raidal}).
We take a simple SUSY grand unified theory
(GUT) model naturally incorporating the
standard SUSY realization \cite{cllsw,dss}
of hybrid inflation \cite{hybrid}, which
does not require tiny parameters and is,
undoubtedly, one of the most promising
inflationary scenarios. (For extensions of
standard SUSY hybrid inflation, see
Ref.~\cite{cairo}.) In global SUSY, the
flatness of the inflationary path at tree
level is guaranteed by a ${\rm U}(1)$
R-symmetry. The $\eta$ problem \cite{cllsw}
of sizable supergravity (SUGRA)
contributions to the inflaton mass on the
inflationary path, which could easily
invalidate inflation, is reduced, in this
case, to controlling the magnitude of a
single term in the K{\" a}hler potential
\cite{lss} (see also Ref.~\cite{nshift}).
Finally, radiative corrections provide
\cite{dss} a logarithmic slope along the
classically flat direction, needed for
driving the inflaton towards the SUSY vacua.

\par
The $\mu$ problem is solved by employing the
mechanism of Ref.~\cite{lr}. (For an
alternative solution of the $\mu$ problem,
see Ref.~\cite{virtualtriplet}.)
The global R-symmetry of the model forbids
the appearance of a $\mu$ term in the
superpotential. On the contrary, it allows
the existence of the trilinear term $Sh_1h_2$,
where $S$ is the gauge singlet inflaton of
the standard SUSY hybrid inflation. After the
GUT gauge symmetry breaking, the soft
SUSY-breaking terms, which generally violate
the R-symmetry, give rise to a suppressed
linear term in $S$ and, thus, this field
acquires a VEV of order the electroweak
scale divided by a small coupling constant.
The above trilinear coupling can then yield a
$\mu$ term of the right magnitude.

\par
The inflaton consists of two complex scalar
fields with tree-level couplings to the
electroweak Higgses and Higgsinos derived
from the above trilinear term. After the end
of inflation, it oscillates about the SUSY
vacuum and eventually decays predominantly
into electroweak Higgs superfields via these
tree-level couplings, thereby reheating the
universe. We find that, in the case of
one fermion family, both heavy RHN and
${\rm SU}(2)_L$ triplet superfields are
necessary if diagrams producing a nonzero
lepton asymmetry are to exist. Since these
heavy fields can only appear in intermediate
states of the inflaton decay, we must create
the asymmetry directly from this decay.
Indeed, leptogenesis can occur in the
subdominant decay of the inflaton into
lepton and Higgs superfields through the
interference between different one-loop
diagrams with RHN and ${\rm SU}(2)_L$
triplet exchange respectively. The lepton
asymmetry is proportional to a novel
CP-violating invariant product of coupling
constants.

\par
For a nonzero asymmetry, we also need to
include some couplings in the superpotential
that explicitly violate both the
${\rm U}(1)$ R-symmetry and its $Z_2$
matter parity subgroup, which remains
unbroken by the soft SUSY-breaking terms.
Although these couplings involve superheavy
fields, the matter parity violation may have
some observable consequences at low energy
such as the possible instability of the
lightest SUSY particle (LSP). Indeed, if
this particle contains a Higgsino component,
it could decay predominantly into a pair of
Higgses and a lepton. This channel can,
though, be easily blocked kinematically if
the LSP is not too heavy (see
Sec.~\ref{sec:rsym}). On the contrary, some
R-parity violating slepton decays which may
be detectable in the future colliders are
typically present (see
Sec.~\ref{sec:numerics}).

\par
We find that the value of the BAU from the
Wilkinson microwave anisotropy probe
(WMAP) data \cite{wmap} can be easily
achieved given constraints from other
observables, notably the reheat temperature
and neutrino masses and mixing, and
CP-violating phases of order unity. However,
the constraint from $\nu$ masses and mixing
is considerably weakened by the existence of
two separate contributions, namely those of
the usual seesaw mechanism and from the
Higgs ${\rm SU}(2)_L$ triplets (see alsoc
Ref.~\cite{Senjanovic}). On the contrary,
the requirement that the initial lepton
asymmetry is protected from lepton number
violating $2\rightarrow 2$ scattering
processes which are in equilibrium in the
early universe imposes very stringent
constraints on the parameters of the model,
which we also take into account in our
analysis. The prediction for the spectral
index of density perturbations is typical of
SUSY hybrid inflation models (see e.g.
Ref.~\cite{lectures}).

\par
Thus, an acceptable value of the BAU can be
obtained within a consistent model of
cosmology and particle physics, without
requiring additional fine-tuned coupling
constants and without necessarily putting
strong constraints on observables such as
neutrino masses and mixing. Moreover,
although the scenario requires violation of
the R-symmetry, it is not necessary to
introduce superpotential terms which would
lead to currently observable R-symmetry
violating effects.

\par
In Sec.~\ref{sec:model}, we introduce our
model and describe some of its salient
features. In Sec.~\ref{sec:invariants}, we
present the CP-violating invariant products
of coupling constants which enter into the
primordial lepton asymmetry, while, in
Sec.~\ref{sec:bau}, we sketch the
calculation of the BAU. The effects of
R-symmetry violation are discussed in
Sec.~\ref{sec:rsym}. The constraints from
neutrino masses and the preservation of
the initial lepton asymmetry are given in
Sec.~\ref{sec:neutrino}, and our numerical
results in Sec.~~\ref{sec:numerics}.
Finally, our conclusions are summarized in
Sec.~\ref{sec:concl}.

\section{The Model}
\label{sec:model}

\par
The model has gauge group ${\rm SU}(3)_c
\times{\rm SU}(2)_L\times{\rm U}(1)_Y\times
{\rm U}(1)_{B-L}$ and a global R-symmetry
${\rm U}(1)_R$, which is though explicitly
broken by some terms in the superpotential
(see below). In addition to the
corresponding vector superfields and the
usual MSSM chiral superfields $h_1$, $h_2$
(Higgs ${\rm SU}(2)_L$ doublets), $l_i$
(${\rm SU}(2)_L$ doublet leptons), $e^c_i$
(${\rm SU}(2)_L$ singlet charged leptons),
$q_i$ (${\rm SU}(2)_L$ doublet quarks),
$u^c_i$, $d^c_i$ (${\rm SU}(2)_L$ singlet
quarks) with $i=1,2,3$ being the family
index, we introduce chiral superfields
$\nu^c_i$ (RHNs), $S$, $\phi$, $\bar{\phi}$
singlets under the SM gauge group and $T$,
$\bar{T}$ in the adjoint representation of
${\rm SU}(2)_L$ with $Y=1,~-1$ respectively.
(As usual, we will use the same symbol to
denote the superfield and its scalar
component, the distinction being clear in
context.) The charges under
${\rm U}(1)_{B-L}$ and ${\rm U}(1)_R$ are
given in Table I.

\begin{table}
\label{table:1}
\caption{${\rm U}(1)$ charges of superfields}
\begin{ruledtabular}
\begin{tabular}{c|ccccccccccccc}
& $S$ & $\phi$ & $\bar{\phi}$ & $T$ & $\bar{T}$
& $h_1$ & $h_2$ & $l$ & $\nu^c$ & $e^c$ & $q$
& $u^c$ & $d^c$ \\ \hline
$B-L$ & 0 & 1 & -1 & 2 & 0 & 0 & 0 & -1 & 1 & 1
& 1/3 & -1/3 & -1/3 \\
$R$ & 2 & 0 & 0 & 0 & 2 & 0 & 0 & 1 & 1 & 1 & 1
& 1 & 1
\end{tabular}
\end{ruledtabular}
\end{table}

The superpotential is
\bea
\label{W}
W &=& \kappa S(\phi\bar{\phi}-M^2)+
\lambda S(h_1h_2) \nonumber \\
&+&h_{eij}(h_1l_i)e^c_j+h_{uij}(h_2q_i)u^c_j+
h_{dij}(h_1q_i)d^c_j \nonumber \\
&+&h_{\nu ij}(h_2l_i)\nu^c_j+h_{Tij}l_i\eps
Tl_j+h_{\bar{T}}h_2\eps\bar{T}h_2 \nonumber \\
&+& (M_{\nu^cij}/M^2)\bar{\phi}^2\nu^c_i\nu^c_j
+ (M_T/M^2)\bar{\phi}^2 T \bar{T} \nonumber \\
&+&(\lambda_i/M_S)\bar{\phi}(h_1 h_2)\nu^c_i
+(\lambda'_i/M_S)\bar{\phi}h_1\eps Tl_i+\cdots,
~~
\eea
where $M$ is a mass parameter of the order of
the GUT scale, $M_S$ is the string scale
$\simeq 5\times 10^{17}~{\rm GeV}$, $\eps$ is
the $2\times 2$ antisymmetric matrix with
$\eps_{12}=1$, $(X Y)$ indicates the
${\rm SU}(2)_L$ invariant product
$\eps_{ab}X_aY_b$ and $T\equiv T_a\sigma_a/
\sqrt{2}$, $\bar{T}\equiv\bar{T}_a\sigma_a/
\sqrt{2}$ with $\sigma_a$ ($a=1,2,3$) being
the Pauli matrices. The ellipsis represents
terms of order higher than four and
summation over indices is implied. The only
${\rm U}(1)_R$ violating terms which we
allow in the superpotential are the two
explicitly displayed terms in the last line
of the right hand side (RHS) of
Eq.~(\ref{W}), which are necessary for
leptogenesis. We can show that baryon number
($B$) is automatically conserved to all
orders as a consequence of ${\rm U}(1)_R$.
The argument goes as in
Ref.~\cite{nonthtripletdec} and is not
affected by the presence of the above
${\rm U}(1)_R$ breaking superpotential
terms. Lepton number ($L$) is then also
conserved as implied by the presence of
${\rm U}(1)_{B-L}$.

\par
The inflationary trajectory is as described
in Ref.~\cite{lr}: for $\kappa<\lambda$, it
is parametrized by $S$, $|S|>S_c=M$, with
all the other fields vanishing and has a
constant energy density $\kappa^2M^4$ at
tree level. Here, the dimensionless
parameters $\kappa$, $\lambda$ and the mass
$M$ are taken real and positive by
redefining the phases of the superfields.
There are radiative corrections \cite{dss}
which lift the flatness of this classically
flat direction leading to slow-roll
inflation until $|S|$ reaches the
instability point at $|S|=M$ as one can
deduce from the $\epsilon$ and $\eta$
criteria \cite{lectures}. The quadrupole
anisotropy of the CMBR and the number of
e-foldings $N_Q\simeq\ln[1.88\times 10^{11}
\kappa^{1/3}(M/{\rm GeV})^{2/3}(T_{\rm rh}/
{\rm GeV})^{1/3}]$ \cite{lectures} of our
present horizon scale during inflation are
given by the Eqs.~(2)-(4) of Ref.~\cite{atmo}
with the two last terms in the RHS of
Eq.~(3) divided by two since the
${\rm SU}(2)_R$ doublet chiral superfields
$l^c$, $\bar l^c$ of this reference are now
replaced by the SM singlets $\phi$,
$\bar\phi$ (see also Ref.~\cite{atmotalk}).

\par
When the value of $|S|$ falls below $M$,
a $B-L$ breaking phase transition occurs
provided that $\kappa<\lambda$. The fields
evolve towards the realistic SUSY minimum
at $\vev{S}=0$, $\vev{\phi}=
\vev{\bar{\phi}}=M$, $\vev{h_1}=\vev{h_2}
=0$, where $\vev{\phi}$, $\vev{\bar{\phi}}$
are taken real and positive by a $B-L$
rotation (there is also an unrealistic
SUSY minimum which is given below).
Actually, with the addition of
soft SUSY-breaking terms, the position of
the vacuum shifts \cite{lr} to nonzero
$\vev{S}\simeq -m_{3/2}/\kappa$, where
$m_{3/2}$ is the gravitino mass, and a small
effective $\mu$ term with $\mu\simeq -
\lambda m_{3/2}/\kappa$ is generated from
the superpotential coupling $\lambda
S(h_1h_2)$. Subsequently, the inflaton
degrees of
freedom $S$ and $\theta\equiv (\delta\phi+
\delta\bar{\phi})/\sqrt{2}$ ($\delta\phi=
\phi-M$, $\delta\bar{\phi}=\bar{\phi}-M$)
with mass $m_{\rm inf}=\sqrt{2}\kappa M$
oscillate about this minimum and decay to
MSSM degrees of freedom reheating the
universe. The predominant decay channels of
$S$ and $\theta$ are to fermionic and
bosonic $h_1$, $h_2$ respectively via
tree-level couplings derived from the
superpotential terms $\lambda S(h_1h_2)$ and
$\kappa S\phi\bar{\phi}$. Note that, if
$\kappa>\lambda$, the system would end up in
the unrealistic SUSY minimum at $\phi=
\bar\phi=0$, $|h_1|=|h_2|\simeq (\kappa/
\lambda)^{1/2}M$, which is degenerate with
the realistic one (up to $m_{3/2}^4$) and is
separated from it by a potential barrier of
order $m_{3/2}^2M^2$.

\par
The RHNs and the ${\rm SU}(2)_L$ triplets
acquire masses $M_{\nu^cij}$ and $M_T$
respectively after the spontaneous breaking
of ${\rm U}(1)_{B-L}$ by $\langle\phi\rangle$,
$\langle\bar{\phi}\rangle$. The terms
which appear in the fourth line of the RHS
of Eq.~(\ref{W}) can also be written in the
form $\lambda_{\nu^cij}\bar{\phi}^2\nu^c_i
\nu^c_j/M_S$, $\lambda_{T}\bar{\phi}^2 T
\bar{T}/M_S$ making it clear that the RHN
and ${\rm SU}(2)_L$ triplet masses are
suppressed by a factor $M/M_S$ relative to
$M$. It is possible to redefine superfields
to obtain effective mass terms $M_{\nu^c_i}
\nu^c_i\nu^c_i$ (which are diagonal in the
flavor space) and $M_TT\bar{T}$ with
$M_{\nu^c_i}$ and $M_T$ real and positive.

\par
Similarly, after the ${\rm U}(1)_{B-L}$
breaking, the explicitly displayed terms
in the last line of the RHS of Eq.~(\ref{W})
which violate ${\rm U}(1)_R$ give rise to
effective $B-L$ and matter parity violating
operators $\zeta_i(h_1h_2)\nu^c_i$ and
$\zeta'_ih_1\eps Tl_i$, where $\zeta_i$ and
$\zeta'_i$ are suppressed by one power of
$M/M_S$. If we require that the magnitude
of the dimensionless coupling constants
$\lambda_i$ and $\lambda'_i$ is less than
unity, we obtain the bound $|\zeta_i|$,
$|\zeta'_i|\leq M/M_S$. Note that, although
$\zeta'_i$ can be made real and positive by
redefining the phase of $l_i$, there is no
phase freedom left which can do the same for
$\zeta_i$. This can be shown by considering
the rephasing invariant $\zeta_i^{*2}\mu^2
M_{\nu_i^c}$ (no summation over $i$) with
$\mu$ and $M_{\nu_i^c}$ already made real.
The coupling constants $\zeta_i$, thus,
remain in general complex. It is, of course,
possible to write down many other R-symmetry
violating operators. However, they are not
necessary for a nonzero lepton asymmetry to
be created.

\par
After electroweak symmetry breaking, the
triplet $T$ also acquires
\cite{triplet,virtualtriplet} a VEV
$\vev{T}=-h_{\bar{T}}\vev{h_2}^2/M_T$. This
is due to the mass term $M_TT\bar{T}$ and
the coupling $h_{\bar{T}}h_2\eps\bar{T}h_2$,
where $h_{\bar{T}}$ is made real and
positive by a
redefinition of the phase of $\bar{T}$ and
a compensating rephasing of $T$ and $l_i$
in order to retain the positivity of $M_T$
and $\zeta'_i$. Note that $\vev{h_1}$,
$\vev{h_2}$ can be taken real because of
the reality of $B\mu\simeq -2\lambda
m_{3/2}^2/\kappa$ \cite{lr}. So $\vev{T}$ is
also real.

\par
Performing appropriate flavor rotations of
the $l_i$ and $e^c_j$, the Yukawa coupling
constant matrix $h_{eij}$ can be
diagonalized with real and positive entries
$h_{ei}$ in the diagonal. The neutrino
components ($\nu_i$) of $l_i$ are then in
the weak interaction basis. (The rephasing
of $l_i$ which was used to make $\zeta'_i$
real and positive should actually be
performed after these rotations and should
be accompanied by a compensating rephasing
of $e^c_i$ so that $h_{ei}$ remains real and
positive.) The coupling of $T$ to a pair of
lepton ${\rm SU}(2)_L$ doublets cannot be
made diagonal since no further flavor
rotations of the $l_i$ are allowed.
Moreover, if we define superfields such that
$M_T$, $\zeta'_i$, $\mu$ and $h_{\bar{T}}$
are real then it is in general not possible
to make the $h_{Tij}$ real since, once $S$,
$\phi$, $\bar{\phi}$ acquire nonzero VEVs,
we have rephasing invariants $M_Th_{Tij}
h_{\bar{T}}^*\mu^2{\zeta'_i}^*{\zeta'_j}^*$
(no summation). Finally, the Yukawa coupling
constants $h_{\nu ij}$ also remain in
general complex as one can easily deduce
from the rephasing invariants
$h_{\bar{T}}^{*2}{\zeta'_i}^{*2}M_T^2\mu^2
M_{\nu^c_j}^*h_{\nu ij}^2$ (no summation
over indices).

\par
The calculation of lepton asymmetry produced
in $S$ and $\theta$ decays is quite
straightforward
but somewhat lengthy, and differs in detail
from the usual case where leptogenesis
occurs via the decay of RHN and/or
${\rm SU}(2)_L$ triplet superfields. Since
we consider the interference of two one-loop
diagrams, we will need to calculate the real
parts of loop integrals which require
renormalization.

\section{CP-violating invariants}
\label{sec:invariants}

\par
To produce a nonvanishing net $B-L$
asymmetry from inflaton decay, the theory
must contain one or more physical
CP-violating quantities. These are products
of coupling constants, corresponding to
operators noninvariant under CP
conjugation, which are nonreal and are not
affected by the redefinition of fields by
complex phases (or other global symmetries).
In this case, we consider the terms of $W$
in Eq.~(\ref{W}). Since leptogenesis takes
place at reheating, we work in the vacuum
where $\vev{\phi}=\vev{\bar{\phi}}=M$. For
simplicity, we ignore the couplings of the
inflaton to RHNs and ${\rm SU}(2)_L$
triplets from the terms in the fourth line
of the RHS of Eq.~(\ref{W}) together with
its couplings resulting from the two
subsequent terms. The inclusion of these
couplings would only complicate the analysis
without altering the character of our
mechanism. So these four superpotential
terms are replaced by the effective mass
terms $M_{\nu^c_i} \nu^c_i\nu^c_i$,
$M_TT\bar{T}$ and couplings
$\zeta_i(h_1h_2)\nu^c_i$ and $\zeta'_ih_1
\eps Tl_i$. The condition of CP invariance
is simply that it should be possible to make
all products of coupling constants real by
field redefinitions.

\par
We consider here only rephasing invariants
which, in contrast to the standard invariant
used in Ref.~\cite{lepto}, would exist even
if there was a single fermion family. We
find the following independent invariants
(no summation over repeated indices):
\bea
\label{invariant}
I_{1ij}&=&M_Th_{\nu ij}\zeta_j(M_{\nu^c_j}
h_{\bar{T}}\zeta'_i)^*, \nonumber \\
I_{2ijk}&=&M_Th_{\nu ij}h_{\nu kj}
(M_{\nu^c_j}h_{Tik}h_{\bar{T}})^*,
\eea
whose imaginary parts violate CP invariance.
Any combination of the two (modulo real
invariants) results in a third invariant
such as $I_{3ijk}=\zeta'_ih_{\nu kj}(\zeta_j
h_{Tik})^*$. In each case, the expression
involves both the ${\rm SU}(2)_L$
triplet and RHN couplings, thus the
generation of a $B-L$ asymmetry is
independent of the sources of CP violation
considered in previous scenarios and we
require novel decay diagrams. The invariants
$I_{1ij}$, $I_{2ijk}$, $I_{3ijk}$ are the
minimal ones in the sense that they have
the smallest possible number of trilinear
superpotential couplings.

\par
It is important to note that the invariant
$I_{1ij}$ can be split in two parts
${\zeta'_i}^*M_Th_{\bar{T}}^*$
and $\zeta_jM_{\nu^c_j}^*h_{\nu ij}$
corresponding to effective operators which
carry (opposite) nonzero $B-L$ charges and
involve only light fields since the heavy
ones can be contracted. These light fields
include bosonic or fermionic $h_1$, $h_2$
and their conjugates. So, $I_{1ij}$ is, in
principle, suitable for leptogenesis which
requires the interference of two $B-L$
violating diagrams. (Recall that the
inflaton field couples at tree level to the
electroweak Higgs superfields $h_1$, $h_2$
and can decay only to light particles.)
This is not the case with the rephasing
invariant $I_{2ijk}$ since there are no
$h_1$'s among the light fields of the
corresponding (effective) operator. On the
contrary, the invariant $I_{3ijk}$ has all
the above good properties of $I_{1ij}$, but
one of the interfering amplitudes turns out
to be zero in this case. The reason is that
it involves a real (with vanishing
absorptive part) on-mass-shell off-diagonal
self-energy between $l$ and $h_1$ which, in
the on-shell (OS) renormalization scheme,
vanishes (see Sec.~\ref{sec:bau}). It is
clear that any invariant which can be useful
for leptogenesis must involve $I_{1ij}$ and,
thus, the effective coupling constants
$\zeta_j$, $\zeta'_i$. So, the explicit
violation of matter parity is essential for
our scheme. This is another novel feature of
this leptogenesis scenario.

\section{Baryon asymmetry}
\label{sec:bau}

\par
\begin{figure}[t]
\includegraphics[width=86mm,angle=0]{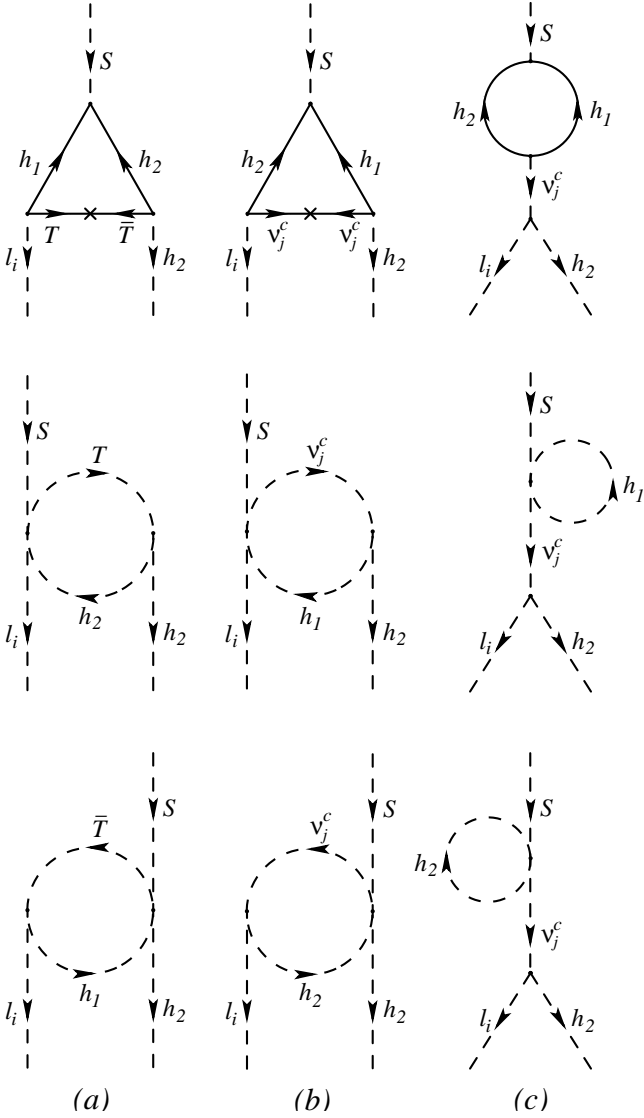}
\caption{\label{fig:1} The nine one-loop
diagrams for the $L$ violating decay
$S\rightarrow\tilde{l}_i\,h_2$. The solid
(dashed) lines represent the fermionic
(bosonic) component of the indicated
superfield. The arrows depict the
chirality of the superfields and the
crosses are mass insertions in fermion
lines.}
\end{figure}
\begin{figure}[t]
\includegraphics[width=86mm,angle=0]{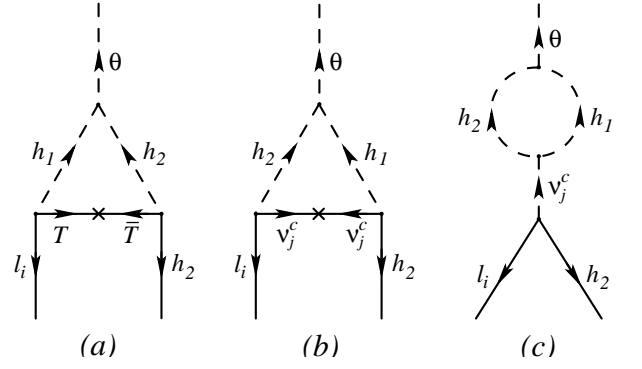}
\caption{\label{fig:2} The three one-loop
diagrams for the $L$ violating decay
$\theta^*\rightarrow l_i\,\tilde{h}_2$.
The notation is as in Fig.~\ref{fig:1}.}
\end{figure}

The CP-violating rephasing invariant
$I_{1ij}$ corresponds to the
product of coupling constants in the
interference of the diagrams in
Fig.~\ref{fig:1}a (Fig.~\ref{fig:2}a) with
the diagrams in Figs.~\ref{fig:1}b and
\ref{fig:1}c (Figs.~\ref{fig:2}b and
\ref{fig:2}c) for the $L$ violating decay of
$S$ ($\theta$). This interference
contributes to the
$L$ asymmetry due to a partial rate
difference in the decays
$S\rightarrow\tilde{l}_i\,h_2$ and $S^*
\rightarrow\tilde{l}_i^*\,h_2^*$ ($\theta^*
\rightarrow l_i\,\tilde{h}_2$ and $\theta
\rightarrow\bar{l}_i\,\bar{\tilde{h}}_2$),
where bar and tilde represent the
antifermion and the SUSY partner
respectively. Both the decaying inflaton
($S$ or $\theta$), which we take at rest,
and the decay products must be on mass
shell. For simplicity, we consider that all
the propagating and external MSSM particles
in the diagrams are massless. Also, we
perform the calculation in the limit of
exact SUSY.

\par
In each case, the resulting contribution
to the $L$ asymmetry is proportional
to both $\mbox{Im}\,I_{1ij}$ and the
imaginary part of the interference of the
relevant `stripped' diagrams with the
dimensionless coupling constants and the
$M_T$, $M_{\nu^c_j}$ mass insertions
factored out (we keep, though, the
$m_{\rm inf}$ factor appearing in the
scalar coupling $\theta^*h_1h_2$).
The `stripped' diagrams in
Figs.~\ref{fig:1}a, \ref{fig:1}b and
\ref{fig:1}c are denoted by $F^S_{ain}$,
$F^S_{bijn}$ and $F^S_{cijn}$ respectively
with $i$ and $j$ being the family indices
and $n=1,2,3$ the serial number of the
diagram. Similarly, the `stripped'
diagrams in Figs.~\ref{fig:2}a, \ref{fig:2}b
and \ref{fig:2}c are $F^\theta_{ai}$,
$F^\theta_{bij}$ and $F^\theta_{cij}$. Thus,
the total net $L$ asymmetries $\eps_{|S}$
and $\eps_{|\theta}$ generated per $S$ and
$\theta$ decay respectively are given by
\bea
\label{epsSTheta}
\eps_{|S}&=&-2\frac{|\lambda|^2}{\Gamma}
\mbox{Im}\,I_{1ij}\mbox{Im}\,[F^S_{ain}
{F^S_{bijn}}^*+F^S_{ain}{F^S_{cijn}}^*],
\nonumber \\
\eps_{|\theta}&=&-2\frac{|\lambda|^2}
{\Gamma}\mbox{Im}\,I_{1ij}\mbox{Im}\,
[F^\theta_{ai}{F^\theta_{bij}}^*+
F^\theta_{ai}{F^\theta_{cij}}^*],
\eea
where $\Gamma=|\lambda|^2 m_{\rm inf}/8\pi$
is the rate of the tree-level decays
$S\rightarrow\bar{\tilde{h}}_1\,
\bar{\tilde{h}}_2$ and $\theta\rightarrow
h_1\,h_2$, and summation over the indices
$i$, $j$, $n$ and integration over the phase
space of the final particles is implied.

\par
We see that the one-loop diagrams for the
$L$ violating decay of the inflaton which
contain ${\rm SU}(2)_L$ triplet exchange
are exclusively of the vertex type (see
Figs.~\ref{fig:1}a and \ref{fig:2}a).
On the contrary, the diagrams with a RHN
exchange are both of the vertex
(Figs.~\ref{fig:1}b and \ref{fig:2}b) and
self-energy (Figs.~\ref{fig:1}c and
\ref{fig:2}c) \cite{covi} type. Each of
the three vertex diagrams in
Fig.~\ref{fig:1}a (Fig.~\ref{fig:1}b)
possesses a logarithmic ultraviolet (UV)
divergence. However, one can easily show
that their sum equals $m_{\rm inf}$ times
the vertex diagram in Fig.~\ref{fig:2}a
(Fig.~\ref{fig:2}b), which is UV finite.
Similarly, one can show that, although each
of the three self-energy diagrams in
Fig.~\ref{fig:1}c has a quadratic
divergence, their sum equals the self-energy
diagram in Fig.~\ref{fig:2}c multiplied by
$m_{\rm inf}$. However, the latter is not UV
finite. It rather possesses a logarithmic
divergence and, thus, needs renormalization
(see below).

\par
The above relations between the diagrams for
the $L$ violating decays of $S$ and $\theta$
imply that $\eps_{|S}=\eps_{|\theta}\equiv
\eps$. We can, thus, concentrate on the
calculation of $\eps_{|\theta}$ which is
simpler. As already mentioned, the vertex
diagrams in Figs.~\ref{fig:2}a and
\ref{fig:2}b are finite (both their real and
imaginary parts) and, thus, well-defined and
independent of the renormalization scheme
used. However, the diagram in
Fig.~\ref{fig:2}c involving a divergent
self-energy loop requires us to apply a
renormalization condition. As shown in
Ref.~\cite{pilaftsis}, the appropriate
renormalization scheme, in this case, is
the on-shell (OS) scheme. In a general
theory with scalars $S_i$, the OS conditions
on the renormalized self-energies
$\hat{\Pi}_{ij} (p^2)$ are as follows:
\beq
\mbox{Re}\,\hat{\Pi}_{ij}(\mu_i^2)=
\mbox{Re}\,\hat{\Pi}_{ij}(\mu_j^2)=0
\eeq
for the off-diagonal self-energies
($i\neq j$) and
\beq
\lim_{p^2\rightarrow \mu_i^2}\frac{1}
{p^2-\mu_i^2}\mbox{Re}\,\hat{\Pi}_{ii}(p^2)
=0
\eeq
for the diagonal ones (see e.g.
Ref.~\cite{Pilaftsisren}). Here we take a
basis where the renormalized mass matrix is
diagonal with eigenvalues $\mu_i$.
The imaginary part of the self-energies
is finite and, thus, not renormalized.
In Fig.~\ref{fig:2}c, we have an
off-diagonal self-energy diagram between the
scalars $\theta$ and $\tilde{\nu}^c_j$.
Given that $\theta$ is on mass shell, the
real part of this diagram vanishes in the OS
scheme. The imaginary part, however, gives a
finite contribution.

\par
The lepton asymmetry per inflaton decay is
calculated by making use of the software
packages of Ref.~\cite{hahn} and is found to
be given by
\beq
\eps=\eps_{VV}+\eps_{VS},
\label{eps}
\eeq
where
\beq
\eps_{VV}=\frac{3}{128\pi^4}\,\frac{
\mbox{Im}\,I_{1ij}}{m_{\rm inf}^2}\,
\mbox{Im}\,[f(y_T)f(y_j)^*]
\label{vv}
\eeq
is the contribution from the interference
of the two vertex diagrams in
Figs.~\ref{fig:2}a and \ref{fig:2}b, and
\beq
\label{vs}
\eps_{VS}=-\frac{3}{64\pi^3}\,\frac{
\mbox{Im}\,I_{1ij}}{m_{\rm inf}^2-
M_{\nu_j^c}^2}\,\mbox{Re}\,f(y_T)
\eeq
is the contribution from the interference
of the vertex and self-energy diagrams in
Figs.~\ref{fig:2}a and \ref{fig:2}c with
$f(y)=\pi^2/6-{\rm Li}_2(1+y+i\varepsilon)$
(${\rm Li}_2$ is the dilogarithm
\cite{dilog}), $y_T=m_{\rm inf}^2/M_T^2$,
$y_j=m_{\rm inf}^2/M_{\nu_j^c}^2$ and
summation over the family indices $i$, $j$
implied. It should
be emphasized that Eq.~(\ref{vs}) holds
\cite{pilaftsis} provided that the decay
width of $\nu^c_j$ is $\ll |m^2_{\rm inf}-
M^2_{\nu^c_j}|/m_{\rm inf}$, which is well
satisfied in our model if $M_{\nu^c_j}$ is
not unnaturally close to $m_{\rm inf}$.

\par
The equilibrium conditions including
nonperturbative electroweak reactions above
the critical temperature of the electroweak
phase transition yield a relation between
the baryon number density $n_B$ and the
$B-L$ number density $n_{B-L}$, which
allows us to find $n_B$ in terms of
the $n_{B-L}$ produced from inflaton decay,
assuming that $B-L$ is conserved at
temperatures well below $M_T$ and
$M_{\nu^c_i}$ (see Sec.~\ref{sec:neutrino}).
In the MSSM with soft SUSY-breaking
terms, we have $n_B/s=(28/79)n_{B-L}/s$
\cite{IbanezQ}. If we imagine the
inflaton to decay instantaneously out of
equilibrium creating initial lepton number
density $n_{L,{\rm init}}$ then
\beq
\frac{n_B}{s}=-\frac{28}{79}\,
\frac{n_{L,{\rm init}}}{s}=-\frac{28}{79}\,
\eps\,\frac{n_{\rm inf}}{s}=-\frac{21}{79}\,
\eps\,\frac{T_{\rm rh}}{m_{\rm inf}}
\label{nb}
\eeq
using the standard relation $n_{\rm inf}/
s\equiv (n_S+n_\theta)/s=3T_{\rm rh}/4
m_{\rm inf}$ for the inflaton number
density. Here $s$ is the entropy
density. The reheat temperature is given
by
\beq
\label{Trh}
T_{\rm rh}=\left(\frac{45}{2\pi^2 g_*}
\right)^{\frac{1}{4}}
(\Gamma m_{\rm P})^{\frac{1}{2}},
\eeq
where $g_*$ counts the relativistic
degrees of freedom taking account of the
spin and statistics and is equal to
$228.75$ for the MSSM spectrum.

\section{Effects of R-symmetry violation}
\label{sec:rsym}

\par
One would expect the explicit violation of
${\rm U}(1)_R$ in the superpotential to
have important consequences for
phenomenology and cosmology. The $Z_2$
subgroup of ${\rm U}(1)_R$, which is left
unbroken by the soft SUSY-breaking terms,
is called matter parity since all the
matter (quark and lepton) superfields change
sign under it. Combined with the $Z_2$
fermion parity (under which all fermions
change sign) yields R-parity, which, if
unbroken, guarantees the stability of the
LSP. In our model, however, matter parity is
violated along with the ${\rm U}(1)_R$ by
the two explicitly displayed terms in the
last line of the RHS of Eq.~(\ref{W}), which
are needed for generating an acceptable BAU.
Thus, R-parity is broken and, consequently,
the LSP generically becomes unstable and
decays rapidly, rendering it unsuitable for
the role of dark matter and leading to
distinctive collider signatures.

\par
In our case, if the LSP contains a Higgsino
component, it could decay into a pair of
electroweak Higgs bosons and a lepton. The
dominant diagrams are constructed from the
${\rm U}(1)_R$ and R-parity violating Yukawa
vertices $\zeta_j(h_1h_2)\nu^c_j$
or $\zeta'_ih_1\eps Tl_i$ with the fermionic
$\nu^c_j$ or $T$ connected to the fermionic
$\nu^c_j$ or $\bar{T}$ of the Yukawa
couplings $h_{\nu ij}(h_2l_i)\nu^c_j$ or
$h_{\bar{T}}h_2\eps\bar{T}h_2$ respectively
via a mass insertion. For the numerical
values of the parameters that we consider
(see Secs.~\ref{sec:neutrino} and
\ref{sec:numerics}), we find that the LSP
life-time can be as low as about
$10^{-1}{\rm sec}$. However, it is easy
to block kinematically the LSP decay by
taking its mass to be smaller than twice
the mass of the lightest Higgs boson, which
is very reasonable. Thus, it is perfectly
possible to rescue the LSP as dark matter
candidate.

\par
Our model could also predict the existence
of other R-parity violating processes at
low energies besides the LSP decay. The
R-parity breaking superpotential couplings
involve at least one superheavy field (a RHN
or ${\rm SU}(2)_L$ triplet). On integrating
out these heavy fields, one generally
obtains effective R-parity violating
operators involving only MSSM fields. These
operators, if they have dimension five or
higher, do not lead to detectable processes
since they are suppressed by some powers of
$M_T$ or $M_{\nu^c_i}$. However, dimension
four operators such as the effective scalar
vertices $h_1h_2h_2^*\tilde{l}_i^*$ or
$h_1\tilde{l}_i\tilde{l}_j^*\tilde{l}_k^*$,
which originate from the superpotential
couplings $\zeta_j(h_1h_2)\nu^c_j$,
$h_{\nu ij}(h_2l_i)\nu^c_j$ or $\zeta'_i
h_1\eps Tl_i$, $h_{Tjk}l_j\eps Tl_k$
respectively, can lead to low-energy
R-parity violating processes which may be
detectable in the future colliders (see
Sec.~\ref{sec:numerics}).

\par
It is well known \cite{turner} that, in any
leptogenesis scenario with RHNs or
${\rm SU}(2)_L$ triplets, it is important to
ensure that the primordial lepton asymmetry
is not erased by lepton number violating
$2\rightarrow 2$ scattering processes such
as $l_i\tilde{l}_j\rightarrow h_2^*
\bar{\tilde{h}}_2$ or $l_i\tilde{h}_2
\rightarrow h_2^*\tilde{l}^*_j$ at all
temperatures between $T_{\rm rh}$ and
about $100~{\rm GeV}$. In our model, due to
the presence of the R-parity violating
superpotential couplings, there exist some
extra processes of this type such as
$\tilde{h}_1\tilde{l}_j\rightarrow h_2^*
\bar{\tilde{h}}_2$ or
$\tilde{h}_1\tilde{h}_2\rightarrow h_2^*
\tilde{l}^*_j$, which are derived from
diagrams similar to the ones mentioned above
for the LSP decay. In addition to all these
processes which correspond to effective
operators of dimension five (or higher), we
also have dimension four R-parity (and
lepton number) violating processes such as
$h_1h_2\rightarrow h_2\tilde{l}_i$ or
$h_1\tilde{l}_i\rightarrow \tilde{l}_j
\tilde{l}_k$, which are derived from the
effective four-scalar vertices in the
previous paragraph.

\par
The initial lepton asymmetry is protected
\cite{IbanezQ} by SUSY at temperatures
between $T_{\rm rh}$ and about
$10^{7}~{\rm GeV}$. For $T\lesssim
10^{7}~{\rm GeV}$, one can show that all
the lepton number violating $2\rightarrow 2$
scattering processes which result from
effective operators of dimension five or
higher are well out of equilibrium for the
values of the parameters used here. The
dimension four lepton number violating
processes, however, are generally in
equilibrium and, thus, special care is
needed in order to retain the initial lepton
asymmetry in our model. We will return to
this issue in the next section.

\par
As already explained, the classical flatness
of the inflationary path in the limit of
global SUSY is ensured, in our model, by a
continuous R-symmetry enforcing a linear
dependence of the superpotential on $S$.
This is retained \cite{lss} even after SUGRA
corrections, given a reasonable assumption
about the K{\" a}hler potential. The
solution \cite{lr} to the $\mu$ problem is
also reliant on the R-symmetry. These
aspects of the model are not affected by the
explicit R-symmetry breaking we consider.

\par
In the model, some R-symmetry violating
couplings are present in the superpotential
and some not. Thanks to the
nonrenormalization property of SUSY, this
situation is stable under radiative
corrections, but one may consider it
unnatural since there is no symmetry to
forbid the terms we set to zero.

\section{Neutrino masses and preservation of
lepton asymmetry}
\label{sec:neutrino}

\par
The two distinct sources of neutrino masses
in the model yield the following mass matrix
of light neutrinos:
\beq
m_{\nu ij}=-\vev{h_2}^2\left[\frac{h_{Tij}}
{h_{\bar{T}}M_T}+h_{\nu ik}
\frac{1}{M_{\nu^c_k}}h_{\nu jk}\right].
\label{numass}
\eeq
The terms in this expression are generally
complex with physical relative phases
between the two contributions. Comparing
this to the combination of parameters
appearing in the lepton asymmetry per
inflaton decay, we see that there is
no obvious correlation between the two.
Hence, we have here more freedom to choose
values for the parameters than in the case
of leptogenesis with a single source of
neutrino mass. However, this freedom is
reduced by requiring that the lepton
asymmetry is preserved until the
electroweak phase transition.

\par
As discussed in Sec.~\ref{sec:rsym}, there
exist unsuppressed dimension four R-parity
and lepton number violating $2\rightarrow 2$
scattering processes which can erase the
primordial lepton asymmetry. They originate
from the effective four-scalar operators
$h_1h_2h_2^*\tilde{l}_i^*$ or
$h_1\tilde{l}_i\tilde{l}_j^*\tilde{l}_k^*$
with coupling constants $\sum_j\zeta_j
h^*_{\nu ij}$ or $\zeta'_ih^*_{Tjk}$
respectively. The former operator violates,
in particular, the $i$-lepton number
($L_i$), i.e. the part of the total lepton
number which corresponds to the $i$th
fermion family. Therefore, in order to
retain at least one part of the lepton
asymmetry, we need to impose the condition
that
\beq
\sum_j\zeta_jh^*_{\nu ij}=0,
\label{noerase}
\eeq
for some value of $i$ \cite{foot}. To
generate a $L_i$ asymmetry, we need to
take a nonzero $\zeta'_i$, as it is obvious
from Eqs.~(\ref{invariant}) and
(\ref{eps})-(\ref{vs}). The operators
$h_1\tilde{l}_i\tilde{l}_j^*\tilde{l}_k^*$
with $j,k\neq i$ or $j=k=i$ will then
violate $L_i$, thereby leading to the
erasure of the primordial $L_i$ asymmetry.
To avoid this, we must take $h_{Tjk}$ to
vanish, unless exactly one of the indices is
equal to $i$, i.e. we allow only $h_{Tik}=
h_{Tki}\neq 0$ for $k\neq i$. Moreover, if
$h_{Tik}\neq 0$ then we require $\zeta'_j=0$
for $j\neq i$ to avoid generating the
$L_i$ violating operator $h_1\tilde{l}_j
\tilde{l}^*_i\tilde{l}^*_k$.

\par
We must further ensure that there are no
processes in equilibrium which violate
$L_i$, while conserving $L$. Such processes
result from the four-scalar operators
$h_2\tilde{l}_kh^*_2\tilde{l}^*_i$
($k\neq i$) which originate from the
superpotential coupling $h_{\nu kj}(h_2l_k)
\nu^c_j$. Thus, we must take
\beq
\sum_jh_{\nu kj}h^*_{\nu ij}=0,
\label{fc}
\eeq
for the value of $i$ which satisfies
Eq.~(\ref{noerase}) and all values of $k$
which are different from this $i$. Also,
the operators $h_1\tilde{l}_ih_1^*
\tilde{l}_j^*$ with $j\neq i$, which
originate from the superpotential coupling
$\zeta'_jh_1\eps Tl_j$, violate $L_i$, but
not $L$. So, we must take $\zeta'_j=0$ for
$j\neq i$. Finally, the operators
$\tilde{l_j}\tilde{l}_k\tilde{l}^*_m
\tilde{l}^*_n$ derived from the
superpotential term $h_{Tjk}l_j\eps Tl_k$
could violate $L_i$. However, as explained
above, $h_{Tjk}$ should vanish unless
exactly one index is equal to $i$. Thus, the
remaining operators automatically conserve
$L_i$. In summary, for an asymmetry to be
generated in $L_i$ and not wiped out
by scattering processes, we require that
$\zeta'_i\neq 0$ and $\zeta'_j=0$ for all
$j\neq i$. Also, $h_{Tjk}=0$ unless exactly
one of $j$ or $k$ is equal to $i$ and
Eqs.~(\ref{noerase}) and (\ref{fc}) should
hold for all $k\neq i$.

\par
We take the neutrino mass ordering
$m_1<m_2<m_3$ and adopt the normal
hierarchical scheme of neutrino masses,
where the solar and atmospheric neutrino
${\rm mass}^2$ differences are identified
with $\delta m_{21}^2$ and $\delta m_{31}^2$
respectively. Analysis \cite{fogli} of the
CHOOZ experiment \cite{chooz} shows that the
mixing angle $\theta_{13}$ can be taken
equal to zero. Moreover, the fact that
$\delta m_{21}^2\ll\delta m_{31}^2$ implies
that, when considering atmospheric neutrino
oscillations, it is a good approximation to
set $\delta m_{21}^2=0$. For simplicity, we
further take $\alpha=\beta$, where $\alpha$,
$\beta$ are the Majorana phases in the
leptonic mixing matrix associated with the
two lighter neutrino mass eigenstates
$\hat{\nu}_1$, $\hat{\nu}_2$ respectively.

\par
Under these circumstances, the weak
interaction eigenstate $\nu_1$ decouples
from the other two $\nu_2$, $\nu_3$ in the
neutrino mass matrix. A simple choice of
parameters which leads to this decoupling
is $h_{\nu 1j}=0$ for $j=2,3$,
$h_{\nu i1}=0$ for $i=2,3$ and $h_{T11}=
h_{T1j}=h_{Tj1}=0$ for $j=2,3$.
Eq.~(\ref{fc}) is then automatically
satisfied for $k=1$ and $i=2,3$, which
means that processes corresponding to the
operators $h_2\tilde{l}_1h^*_2\tilde{l}^*_i$
($i=2,3$) which violate $L_1$, while
conserving $L$, are avoided. If we further
take $\zeta'_1=0$, $L_1$ violating and $L$
conserving processes from the operators
$h_1\tilde{l}_1h_1^*\tilde{l}_j^*$ ($j=2,3$)
are also absent. Moreover, our choice of
$h_{Tjk}$ ensures that no such processes
from operators of the type $\tilde{l}_j
\tilde{l}_k\tilde{l}_m^*\tilde{l}_n^*$
appear. Eq.~(\ref{noerase}) with $i=1$
($i=2,3$) receives contribution only from
the term(s) with $j=1$ ($j=2,3$). Also,
$L$ and $L_1$ violating processes from the
operators $h_1\tilde{l}_i\tilde{l}_j^*
\tilde{l}_k^*$ do not exist for our choice
of parameters. Note that there are no $L_1$
violating processes in equilibrium from the
superpotential terms $\zeta'_ih_1\eps Tl_i$,
$h_{Tjk}l_j\eps Tl_k$. The only possible
$L_1$ violating processes come from the
operator $h_1h_2h_2^*\tilde{l}_1^*$, which
is though $L_2$ and $L_3$ conserving.
The fact that $\zeta'_1=0$ implies that no
primordial $L_1$ asymmetry is generated as
one can see from Eqs.~(\ref{invariant}) and
(\ref{eps})-(\ref{vs}). Also, the $L_2$
and $L_3$ asymmetries cannot turn into $L_1$
asymmetry by $2\rightarrow 2$ scattering
processes since, as explained, there are no
such processes in equilibrium which
simultaneously violate $L_1$ and $L_2$ or
$L_3$. All these facts allow us to ignore
the first family and concentrate on the two
heaviest families.

\par
From the recent global analysis \cite{flmm}
of neutrino oscillation data, we take the
best-fit value of the atmospheric neutrino
${\rm mass}^2$ difference which, in the case
of the normal hierarchical scheme, yields
the value $m_3\simeq 5.1\times 10^{-2}~
{\rm eV}$ for the heaviest neutrino mass.
Also, the mixing angle $\theta_{23}$ is
identified with its best-fit value which is
about $\pi/4$ and $m_2$ is taken equal to
zero consistently with the fact that we are
considering only atmospheric neutrino
oscillations. So, the neutrino mass matrix
elements in Eq.~(\ref{numass}) must satisfy
the following restrictions:
\beq
m_{\nu 22}=m_{\nu 33}=m_{\nu 23}=
\frac{1}{2}m_3.
\label{3cond}
\eeq
(Note that, in the case of three neutrino
flavors, our choice of parameters
supplemented, for consistency, with a
vanishing $m_1$ yields the extra restriction
$h_{\nu 11}=0$.) These equalities need not
be exact, however, we will take them as so
for simplicity. We choose to maintain the
$L_3$ asymmetry, which requires that
Eq.~(\ref{noerase}) holds for $i=3$ and
Eq.~(\ref{fc}) for $k=2$, $i=3$. So, we
obtain the following extra conditions:
\bea
h_{\nu 32}^*+Rh_{\nu 33}^*=0,~~~~
\nonumber \\
h_{\nu 22}h_{\nu 32}^*+h_{\nu 23}
h_{\nu 33}^*=0,
\label{2cond}
\eea
where $R=\zeta_3/\zeta_2$. Applying the
results of the previous discussion, we
require $\zeta'_3$ to be nonzero and
$\zeta'_2=0$ (so no $L_2$ asymmetry is
generated); we have also $h_{T22}=h_{T33}=
0$, so the only nonzero coupling constant of
the type $h_{Tjk}$ is $h_{T23}=h_{T32}$.
Given the value of the right handed neutrino
masses, $M_T$ and $h_{\bar{T}}$, the five
conditions in Eqs.~(\ref{3cond}) and
(\ref{2cond}) can be solved, using the mass
formula in Eq.~(\ref{numass}), to find a
relation between the complex Yukawa coupling
constants $h_{ij}$ ($i,\,j=2,\,3$) and the
complex parameters $R$ and $h_{T23}$.

\par
Note that the presence of $L$ and $B-L$
violating $2\rightarrow 2$ scattering
processes leads to a moderate modification
of the numerical factors in Eq.~(\ref{nb}).
However, this modification depends on
details and we will, thus, ignore it.

\section{Numerical results}
\label{sec:numerics}

\par
We saturate the gravitino bound on the
reheat temperature by taking $T_{\rm rh}=
10^{10}~{\rm GeV}$ which is acceptable
\cite{gravitino} provided that the
branching ratio of the gravitino decay to
photons and photinos is less than unity. We
also fix the parameter $\kappa$ to the value
$10^{-4}$. The cosmic microwave background
explorer (COBE) value of the quadrupole
anisotropy of the CMBR ($(\delta T/T)_Q
\simeq 6.6\times 10^{-6}$) \cite{cobe} is
then reproduced for $\lambda\simeq 1.44
\times 10^{-4}$ ($>\kappa$ as it should)
and $M\simeq 3.53\times 10^{15}~{\rm GeV}$.
Thus, $m_{\rm inf}\simeq 4.99\times
10^{11}~{\rm GeV}$. The spectral index of
density perturbations comes out practically
equal to unity.

\par
We now wish to demonstrate that our
leptogenesis mechanism can easily reproduce
the best-fit value of the BAU, $n_B/s\simeq
8.66\times 10^{-11}$, derived from the
recent WMAP data \cite{wmap} with a natural
choice of parameters. For simplicity, we
approximate all the right handed neutrino
masses to the same value $M_{\nu^c}$. (Note
that we do not make use of resonance effects
for degenerate heavy neutrinos, so this
approximation will not strongly influence
our results.) Eq.~(\ref{2cond}) implies that
$h_{\nu 32}=-R^*h_{\nu 33}$ and $h_{\nu 23}=
R h_{\nu 22}$. (Note that $h_{\nu 33}
\neq 0$ for a nonvanishing $L_3$ asymmetry
to be produced.) Substituting these in
Eq.~(\ref{3cond}), we then obtain
\beq
h_{\nu 22}=\pm i\left(\frac{m_3M_{\nu^c}}
{2\vev{h_2}^2}\right)^{\frac{1}{2}}
\frac{1}{(1+R^2)^{\frac{1}{2}}},
\label{h22}
\eeq
\beq
h_{\nu 33}=\pm i\left(\frac{m_3M_{\nu^c}}
{2\vev{h_2}^2}\right)^{\frac{1}{2}}
\frac{1}{(1+R^{*2})^{\frac{1}{2}}},
\label{h33}
\eeq
and
\beq
\pm\frac{R-R^*}
{(1+R^{2})^{\frac{1}{2}}
(1+R^{*2})^{\frac{1}{2}}}=1+
\frac{2h_{T23}\vev{h_2}^2}
{h_{\bar{T}}m_3M_T}.
\label{hT}
\eeq
Note that the signs of $h_{\nu 22}$ and
$h_{\nu 33}$ can be chosen independently;
however, the sign on the left hand side
(LHS) of Eq.~(\ref{hT}) is the product of
the two.

\par
The LHS of Eq.~(\ref{hT}) is
imaginary. Therefore, the real part of the
RHS of this equation must vanish, which
gives
\beq
\mbox{Re}\,h_{T23}=-\frac{h_{\bar{T}}m_3M_T}
{2\vev{h_2}^2}.
\eeq
This implies that
\beq
\frac{R-R^*}{(1+
R^{2})^{\frac{1}{2}}(1+
R^{*2})^{\frac{1}{2}}}=i
\frac{2\mbox{Im}\,(h_{T23})\vev{h_2}^2}
{h_{\bar{T}}m_3M_T}\equiv i\xi,
\label{xi}
\eeq
which yields
\beq
\xi^2|R|^4+2[(1+\xi^2)\cos 2\varphi-1]
|R|^2+\xi^2=0,
\eeq
where $\varphi$ is the phase of the complex
parameter $R$. This equation admits real
solutions for $|R|^2$ provided that
$-1\leq\cos 2\varphi\leq (1-\xi^2)/
(1+\xi^2)$. They turn out to be nonnegative
and are given by
\bea
|R|^2&=&\xi^{-2}\{[1-(1+\xi^2)\cos
2\varphi]\pm (1+\xi^2)^{\frac{1}{2}}
\nonumber \\
& &[(1+\xi^2)\cos^2 2\varphi-2\cos 2\varphi
+(1-\xi^2)]^\frac{1}{2}\}.~~~~~~
\label{zeta}
\eea

\par
From Eqs.~(\ref{invariant}) and
(\ref{eps})-(\ref{nb}), we see that, in our
case, the baryon asymmetry is proportional
to $\zeta'_3h_{\bar{T}}|\zeta_2h_{\nu 33}
\mbox{Im}\,R|$ for given $M_{\nu^c},M_T\geq
m_{\rm inf}/2\simeq 2.5\times
10^{11}~{\rm GeV}$. Using Eqs.~(\ref{h33})
and (\ref{xi}), we can further show that
this expression is proportional to $\zeta'_3
h_{\bar{T}}|\zeta_2\xi^{1/2}R^{1/2}|
(1-\cos 2\varphi)^{1/4}$. As discussed in
Sec.~\ref{sec:model}, the parameters
$\zeta'_3$, $|\zeta_2|$ and $|\zeta_3|=|R|
|\zeta_2|$ should not exceed $M/M_S\simeq
7\times 10^{-3}$. To maximize $n_B/s$, we
saturate this limit on $\zeta'_3$ and on
$|\zeta_2|$ or $|\zeta_3|$ for $|R|\leq 1$
or $|R|>1$ respectively.

\par
In the former case, the baryon asymmetry,
which is $\propto h_{\bar{T}}|\xi|^{1/2}
|R|^{1/2}(1-\cos 2\varphi)^{1/4}$, can be
further maximized by choosing the positive
sign in the RHS of Eq.~(\ref{zeta}) and
taking $\cos 2\varphi=-1$. We then get $|R|=
(1+\xi^{-2})^{1/2}+|\xi|^{-1}$, which is
always $>1$. Our requirement that
$|R|$ does not exceed unity is fulfilled
only asymptotically, i.e. for $|\xi|
\rightarrow\infty$, where $|R|\rightarrow 1$.
The baryon asymmetry is then $\propto
h_{\bar{T}}|\xi|^{1/2}$ and is maximized by
maximizing $h_{\bar{T}}$ and $|\mbox{Im}\,
h_{T23}|$. It would thus be desirable to put
both these quantities equal to unity. The
parameter $h_{\bar{T}}$ can be readily fixed
to unity and this improves the naturalness
of our scheme.
However, if $|\mbox{Im}\, h_{T23}|$ is too
large, some moderate cancellations between
different contributions to the neutrino mass
matrix are required. So, we take values of
$|\mbox{Im}\, h_{T23}|$ which are smaller
than unity, but greater than, say, $10^{-2}$
to be consistent with the requirement that
$|\xi|\gg 1$ for $M_T$ not much bigger than
$m_{\rm inf}/2$.

\par
In the case where $|R|>1$, we should
saturate the limit on $|\zeta_3|$ rather
than the limit on $|\zeta_2|$, which yields
that the baryon asymmetry is $\propto
h_{\bar{T}}|\xi|^{1/2}|R|^{-1/2}
(1-\cos 2\varphi)^{1/4}$. This is maximized
by choosing the negative sign in the RHS of
Eq.~(\ref{zeta}) and taking again $\cos
2\varphi=-1$, which gives $|R|=
(1+\xi^{-2})^{1/2}-|\xi|^{-1}<1$. So,
marginal consistency is achieved again for
$|\xi|\gg 1$, where $|R|\rightarrow 1$.
The baryon asymmetry is again $\propto
h_{\bar{T}}|\xi|^{1/2}$ and is maximized by
taking $h_{\bar{T}}=|\mbox{Im}\,
h_{T23}|=1$. As in the previous case, we
take $h_{\bar{T}}=1$ and $|\mbox{Im}\,
h_{T23}|\sim 10^{-2}$. The result is
obviously the same in both cases to a good
approximation. The phase of $\zeta_2$ is
appropriately adjusted in each case so that
we obtain the maximal baryon asymmetry.

\begin{figure}[t]
\includegraphics[width=84mm,angle=-90]{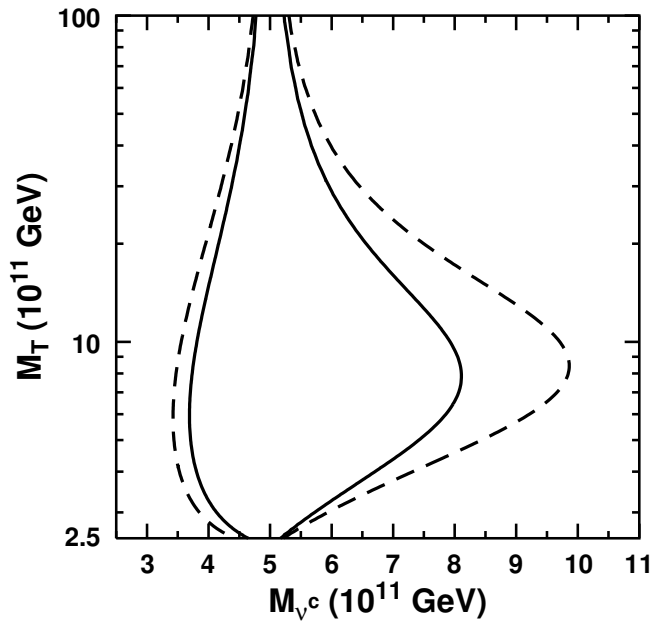}
\caption{\label{fig:3} The solutions of $n_B
/s=8.66\times 10^{-11}$ in the $M_{\nu^c}-
M_T$ plane for $\mbox{Im}\, h_{T23}=0.02$
(solid lines) or $0.04$ (dashed lines). The
values of all the other parameters are given
in the text.}
\end{figure}

\par
For each $M_{\nu^c}, M_T\geq m_{\rm inf}/2$,
we calculate the BAU using
Eqs.~(\ref{eps})-(\ref{nb}) with $\mbox{Im}
\, h_{T23}=0.02$ or $0.04$ and values for
the other parameters as explained above.
We then compare the result with the best-fit
value of the BAU from WMAP. The resulting
solutions in the $M_{\nu^c}-M_T$ plane are
shown in Fig.~\ref{fig:3} with the solid or
dashed lines corresponding to $\mbox{Im}\,
h_{T23}=0.02$ or $0.04$ respectively. Note
that there are two branches in each case,
one with $M_{\nu^c}>m_{\rm inf}$ and one
with $M_{\nu^c}<m_{\rm inf}$. They
correspond to the two possible signs of the
denominator in the RHS of Eq.~(\ref{vs}) for
$\eps_{VS}$. Actually, this quantity, as it
involves a s-channel exchange of a $\nu^c$
boson, can be easily enhanced by letting
$m_{\rm inf}$ approach the $\nu^c$-pole.
This fact assists us to achieve the WMAP
value of $n_B/s$ with natural values of
the parameters. It is important though to
stress that there is no need to get too
near the $\nu^c$-pole for reasonable values
of $M_T$. In fact, the solution is very far
from being unnaturally close to this pole.

\par
As mentioned in Sec.~\ref{sec:rsym}, the
explicit R-parity violation in our model,
which is required for leptogenesis, has some
low-energy signatures which come from
dimension four effective scalar vertices and
may be observable in the future colliders.
Such signatures may typically be the
three-body slepton decay processes
\beq
\tilde{l}_2\rightarrow h_1h_2h_2^*~~
{\rm and}~~
\tilde{l}_2\rightarrow h_1\tilde{l}_3
\tilde{l}_3^*,
\label{decay}
\eeq
which can easily be kinematically allowed.
(Note that, for our choice of parameters,
similar $\tilde{l}_1$ decay processes do
not appear, although they may be present in
the general case.) The effective coupling
constants of the processes in
Eq.~(\ref{decay}) are $\zeta_2h_{\nu 22}^*+
\zeta_3h_{\nu 23}^*$ and $\zeta'_3h_{T23}^*$
respectively. Using the relation $h_{\nu 23}
=Rh_{\nu 22}$ and substituting $h_{\nu 22}$
from Eq.~(\ref{h22}), we can easily show
that the magnitude of the former coupling
constant becomes $\simeq|\zeta_2|
|2\mbox{Im}\,h_{T23}|^{1/2}
(M_{\nu^c}/M_T)^{1/2}$ in the large $|\xi|$
limit. For $\mbox{Im}\,h_{T23}=0.02$ and
taking $M_{\nu^c}\simeq M_T$, which holds
near the right corner of the solid line
in Fig.~\ref{fig:3}, we find that the
magnitude of the effective coupling
constants of the processes in
Eq.~(\ref{decay}) is about
$1.4\times 10^{-3}$ and $1.4\times 10^{-4}$
respectively. The corresponding decay rates
are then of order $10^{-8}~{\rm GeV}$
and $10^{-10}~{\rm GeV}$ respectively for
mass of the decaying slepton $\sim 1~
{\rm TeV}$ and assuming that there is an
appreciable gap between this mass and the
sum of the masses of the decay products.

\par
It is finally interesting to point out that,
as one can readily show, our scheme fails
to provide any useful predictions for the so
far undetermined or not so accurately
determined parameters in the neutrino mixing
matrix, i.e. the three CP-violating phases
and the mixing angle $\theta_{13}$. The
reason is that the number of parameters
is such that successful leptogenesis is
possible whatever the values of the complex
phases and $\theta_{13}$.

\section{Conclusions}
\label{sec:concl}

\par
We proposed a scenario of nonthermal
leptogenesis following supersymmetric hybrid
inflation, in the case where the decay of
the inflaton to both heavy right handed
neutrino and ${\rm SU}(2)_L$ triplet
superfields is kinematically blocked. The
primordial lepton asymmetry is generated
through the direct decay of the inflaton
into light particles. We implemented our
scenario in the context of a simple SUSY
GUT model which incorporates the standard
version of SUSY hybrid inflation. The $\mu$
problem is solved via a ${\rm U}(1)$
R-symmetry which forbids the existence of an
explicit $\mu$ term, while allows a
trilinear superpotential coupling of the
gauge singlet inflaton to the electroweak
Higgs superfields. After the spontaneous
breaking of the GUT gauge symmetry, this
singlet inflaton acquires a suppressed VEV
due to the soft SUSY-breaking terms. Its
trilinear coupling to the Higgs superfields
then yields a $\mu$ term of the right
magnitude.

\par
The main decay mode of the inflaton is to a
pair of electroweak Higgs superfields via
the same trilinear coupling. The initial
lepton asymmetry is created in the
subdominant decay of the inflaton to a
lepton and an electroweak Higgs superfield
via the interference of one-loop diagrams
with right handed neutrino and
${\rm SU}(2)_L$ triplet exchange
respectively. The existence of these
diagrams requires the presence of some
specific superpotential couplings which
explicitly violate the ${\rm U}(1)$
R-symmetry and R-parity. However, the broken
R-parity need not have currently observable
low-energy signatures, although it may have
signatures detectable in future colliders.
Also, the LSP can be made stable and, thus,
be a possible candidate for cold dark
matter.

\par
In our analysis, we took into account the
constraints from neutrino masses and mixing.
There exist, in our model, two separate
contributions to the neutrino mass matrix,
which originate from the usual seesaw
mechanism and from the ${\rm SU}(2)_L$
triplet superfields. The constraints arising
from neutrino masses and mixing alone are
not very stringent. However, the requirement
that the primordial lepton asymmetry not be
erased by lepton number violating processes
before the electroweak phase transition is a
much more stringent constraint on the
parameters of the theory. Taking into
account these constraints we found that the
best-fit value of the BAU from the recent
WMAP data can be easily achieved with
natural values of parameters.

\section*{Acknowledgements}
\par
We thank A. Pilaftsis and K. Tobe for useful
discussions and T. Hahn for his help with
the software packages of Ref.~\cite{hahn}.
T. D. thanks the Michigan Centre for
Theoretical Physics for hospitality and a
pleasant environment during their
Baryogenesis Workshop. This work was
supported by the European Union under RTN
contracts HPRN-CT-2000-00148 and
HPRN-CT-2000-00152.

\def\ijmp#1#2#3{{Int. Jour. Mod. Phys.}
{\bf #1},~#3~(#2)}
\def\plb#1#2#3{{Phys. Lett. B }{\bf #1},~#3~(#2)}
\def\zpc#1#2#3{{Z. Phys. C }{\bf #1},~#3~(#2)}
\def\prl#1#2#3{{Phys. Rev. Lett.}
{\bf #1},~#3~(#2)}
\def\rmp#1#2#3{{Rev. Mod. Phys.}
{\bf #1},~#3~(#2)}
\def\prep#1#2#3{{Phys. Rep. }{\bf #1},~#3~(#2)}
\def\prd#1#2#3{{Phys. Rev. D }{\bf #1},~#3~(#2)}
\def\npb#1#2#3{{Nucl. Phys. }{\bf B#1},~#3~(#2)}
\def\npps#1#2#3{{Nucl. Phys. B (Proc. Sup.)}
{\bf #1},~#3~(#2)}
\def\mpl#1#2#3{{Mod. Phys. Lett.}
{\bf #1},~#3~(#2)}
\def\arnps#1#2#3{{Annu. Rev. Nucl. Part. Sci.}
{\bf #1},~#3~(#2)}
\def\sjnp#1#2#3{{Sov. J. Nucl. Phys.}
{\bf #1},~#3~(#2)}
\def\jetp#1#2#3{{JETP Lett. }{\bf #1},~#3~(#2)}
\def\app#1#2#3{{Acta Phys. Polon.}
{\bf #1},~#3~(#2)}
\def\rnc#1#2#3{{Riv. Nuovo Cim.}
{\bf #1},~#3~(#2)}
\def\ap#1#2#3{{Ann. Phys. }{\bf #1},~#3~(#2)}
\def\ptp#1#2#3{{Prog. Theor. Phys.}
{\bf #1},~#3~(#2)}
\def\apjl#1#2#3{{Astrophys. J. Lett.}
{\bf #1},~#3~(#2)}
\def\n#1#2#3{{Nature }{\bf #1},~#3~(#2)}
\def\apj#1#2#3{{Astrophys. J.}
{\bf #1},~#3~(#2)}
\def\anj#1#2#3{{Astron. J. }{\bf #1},~#3~(#2)}
\def\apjs#1#2#3{{Astrophys. J. Suppl. }
{\bf #1},~#3~(#2)}
\def\mnras#1#2#3{{MNRAS }{\bf #1},~#3~(#2)}
\def\grg#1#2#3{{Gen. Rel. Grav.}
{\bf #1},~#3~(#2)}
\def\s#1#2#3{{Science }{\bf #1},~#3~(#2)}
\def\baas#1#2#3{{Bull. Am. Astron. Soc.}
{\bf #1},~#3~(#2)}
\def\ibid#1#2#3{{\it ibid. }{\bf #1},~#3~(#2)}
\def\cpc#1#2#3{{Comput. Phys. Commun.}
{\bf #1},~#3~(#2)}
\def\astp#1#2#3{{Astropart. Phys.}
{\bf #1},~#3~(#2)}
\def\epjc#1#2#3{{Eur. Phys. J. C}
{\bf #1},~#3~(#2)}
\def\nima#1#2#3{{Nucl. Instrum. Meth. A}
{\bf #1},~#3~(#2)}
\def\jhep#1#2#3{{J. High Energy Phys.}
{\bf #1},~#3~(#2)}
\def\lnp#1#2#3{{Lect. Notes Phys. }
{\bf #1},~#3~(#2)}
\def\appb#1#2#3{{Acta Phys. Polon. B}
{\bf #1},~#3~(#2)}


\begin{thebibliography}{99}
\bibitem{foot}
It is not necessary for this equation, or
the other constraints on $L_i$ violating
couplings, to be satisfied exactly: the
dangerous dimension four operators will not
come into equilibrium if their coupling
constants are less than $10^{-7}$. However,
we take them to vanish exactly in order
to simplify the discussion.
\end{thebibliography}
\end{document}